\documentclass[12pt]{article}
\newlength{\mytopmargin}
\newlength{\myleftmargin}
\newtheorem{theorem}{Theorem}[section]
\newtheorem{proposition}[theorem]{Proposition}
\newtheorem{cor}[theorem]{Corollary}

\usepackage{amsmath,amsfonts,amssymb}

\usepackage{graphicx}

\begin{document}
\title{Fluctuation universality for a class of directed solid-on-solid models}
\author{Benjamin J. Fleming and Peter J. Forrester}
\date{}
\maketitle

\noindent
\thanks{\small Department of Mathematics and Statistics, 
The University of Melbourne,
Victoria 3010, Australia email: {bfleming@ms.unimelb.edu.au \: p.forrester@ms.unimelb.edu.au} 
}

\begin{abstract}
\noindent  Our interest is in a class of directed solid-on-solid models, which may be regarded as  continuum versions of boxed
plane partitions. In the case that the heights are chosen from a uniform distribution, the joint PDF of the heights is the same as
that for the positions in a  finitized bead process recently introduced by the authors and Nordenstam. We use knowledge
of the correlation functions for the latter to show that upon a certain scaling the fluctuations of the heights along the back
row of the solid-on-solid model are given by the Airy process from random matrix theory, as is the case for boxed plane partitions. Moreover, we show that
this limiting distribution remains true if instead of the uniform distribution, the heights are sampled from a general
absolutely continuous distribution.\end{abstract}

\section{Introduction}
For more than a decade now, it has been demonstrated by numerous examples that fluctuation
formulas relating to the largest eigenvalues of certain random matrix ensembles are
also characteristic of fluctuation formulas for certain random growth models (for a recent review
see \cite{FS10}). Because the precise functional form of the random matrix quantities are known
(see e.g.~\cite{Fo10}), there are thus precise theoretical predictions for the functional form of the
corresponding quantities in the growth models. Very recently, the validity of these theoretical
predictions \cite{PS01} has been demonstrated by a high precision experiment involving
curved interface fluctuations of a growing droplet \cite{TS10}.

Crucial to the theoretical predictions is the hypothesis of universality of the fluctuation formulas: they depend on some global symmetry properties of the interface (in particular, is it curved or flat?), but not the details of the microscopic interactions. This hypothesis is generally unproven, as the results for many of the random growth models rely on the integrability of the particular microscopic choice of parameters. One recent advance has been to show that the random matrix fluctuation formulas
predicted on the basis of exact calculation have been shown to hold true for the continuum KPZ equation with narrow wedge initial conditions \cite{ACQ10,SS10}. But mostly universality results are out of reach. 

An outstanding problem of this type relates to last passage percolation. Let $\{a_{ij}\}$ be
independent random variables drawn from a probability distribution with unit mean and finite
variance. The random variable
\begin{equation}\label{ur}
l_N = {\rm max} \sum_{(1,1) \, {\rm u/r} \, (N,N)} a_{ij}
\end{equation}
finishing at $(N,N)$ and forming an up/right directed path when plotted on the integer grid,
is referred to as the last passage time. In the case that each $a_{ij}$ is drawn from an
exponential distribution (or more generally a geometric distribution) the scaled fluctuation of
$l_N$ about its mean $4N$ is known to be identical to the fluctuation of the scaled largest
eigenvalue of a random complex Hermitian matrix \cite{Jo99a}. But for the $a_{ij}$ with
distributions different from these solvable cases, the corresponding theorem is not known.

In this paper we will demonstrate universality  for fluctuations of a certain two-dimensional
directed solid-on-solid model, defined in terms of a grid of height variables $[x_{ij}]_{i,j=1,\dots,N}$.
Each height variable is chosen independently from the same non-negative,  absolutely continuous distribution $\mathcal D$, subject to a constraint on its value relative to its neighbours.
The solid-on-solid  model is defined in Section 2. In Section 3, in the particular case $\mathcal D =
U_{[0,1]}$ (the uniform distribution on $[0,1]$), use is made of a correspondence between this
instance of the solid-on-solid model and the recently introduced finitized bead process
\cite{FFN08}, to obtain a Fredholm determinant formula for the distribution of the largest heights
along the back row of the former. In Section 4
the explicit form of the distribution function between scaled heights on the back row ${\rm O}(N^{2/3})$ away from each other is computed.
The functional form found is precisely that for the Airy process \cite{Ma94,FNH99,PS01},
which in turn is identical to the correlation functions for the GUE minor process at the soft edge
\cite{FN08}. We show in Section 5 that, upon an appropriate choice of scaled variables, this
same correlation function persists for the $x_{ij}$ chosen from a general absolutely continuous distribution  $\mathcal D$,
thereby establishing the claimed universality.

\section{Definition of the model}
\setcounter{equation}{0}
Consider the $N \times N$ integer grid $\{(i,j): 1 \le i,j\le N \}$. At each site $(i,j)$, associate a
height variable $x_{i,j} \mathop{=}\limits^{\rm d} \mathcal D$, where $\mathcal D$ is
an absolutely continuous distribution with density $h(x)$. Furthermore, require that for fixed $i$
$$
x_{i,1} < x_{i,2} < \cdots < x_{i,N}
$$
(heights increase along rows) and that for fixed $j$
$$
x_{1,j} < x_{2,j} < \cdots < x_{N,j}
$$
(heights increase up columns).
Thus a directed up/right lattice path (recall text below (\ref{ur})) must encounter successively larger heights.

Rotate the square grid by 45${}^\circ$ anti-clockwise and mark lines parallel to the $y$-axis through the
lattice points. For an $N \times N$ grid there are $2N - 1$
lines, and on the $l$-th line there are $l$ grid points for $l=1,\dots,N$, and
$2N - l$ grid point for $l=N+1,\dots,2N-1$.  Let $y_i^{(l)}$ denote the height of the $i$-th lattice
point counting from the top on line $j$. We have that $1 \le i \le N(l)$, where
$N(l):= l$, $(i=1,\dots,N)$, and $N(l) := 2N - l$ $(l=N+1,\dots,2N)$
 in keeping with the total
number of particles on line $j$, the relationship
\begin{equation}
x_{jk} = \left \{
\begin{array}{ll} y^{(N+1-j+k)}_{(N+1-j+k) + 1 - k}, & j \ge k \nonumber \\[1.5mm]
y^{(N+1-j+k)}_{(N+1-j+k) + 1 - j}, & j < k,
\end{array} \right.
\end{equation}
and most importantly the interlacings
\begin{equation}\label{7.1}
y_{j+1}^{(l)} < y_j^{(l-1)} < y_{j}^{(l)}, \qquad 1 \le j \le (l-1) \le N
\end{equation}
and
\begin{equation}\label{7.2}
y_{j+1}^{(l)} < y_j^{(l+1)} < y_{j}^{(l)}, \qquad 1 \le j \le 2N - (l+1) \le (N-1).
\end{equation}

Let us denote the interlacings (\ref{7.1}) and (\ref{7.2}) by $Y(\{y\})$. The joint PDF for a
configuration specified by the height coordinates $\cup_{l=1}^{2N} \{ y_j^{(l)} \}_{j=1,\dots,N(l)}$
will then be give in terms of the PDF for the heights $h(x)$ by
\begin{equation}\label{8}
{1 \over C} \prod_{l=1}^{2N} \prod_{j=1}^{N(l)} h(y_j^{(l)}) \chi_{Y(\{ y \})},
\end{equation}
where $\chi_A = 1$ for $A$ true and $\chi_A = 0$ otherwise.

As our first exact result in relation to the joint PDF (\ref{8}), let us compute the normalization $C$.

\begin{proposition}\label{p5}
The normalization $C$ in (\ref{8}) is independent of the PDF $h$, and is given by
\begin{equation}\label{C}
C = \prod_{j=0}^N {\Gamma(1 + j) \over \Gamma(N + j + 1) }.
\end{equation}
\end{proposition}

\noindent
Proof. \quad The PDF $h(x)$ maps $\mathbb R_{\ge 0}$ to $\mathbb R_{\ge 0}$. Its antiderivative
$$
H(x) = \int_0^x h(t) \, dt
$$
is therefore a non-decreasing function from $\mathbb R_{\ge 0}$ to $[0,1]$.Thus if we change variables 
\begin{equation}\label{uH}
u_j^{(l)} = H(y_j^{(l)})
\end{equation}
 we see that the interlacings (\ref{7.1}), (\ref{7.2})
remain true in the variables $\{ u_j^{(l)}  \}$. Furthermore
$$
{d u_j^{(l)}  \over d y_j^{(l)} } = h(y_j^{(l)})
$$
and so
\begin{equation}\label{9}
\prod_{l=1}^{2N} \prod_{j=1}^{N(l)} h(y_j^{(l)}) \chi_{Y(\{y\})} \, d \vec{y}
= \prod_{l=1}^{2N} \prod_{j=1}^{N(l)} U_{[0,1]}(u_j^{(l)})  \chi_{Y(\{u\})}  \, d \vec{u}.
\end{equation}

The measure on the RHS of (\ref{9}) is an example (the case $p=q=N$) of the finitization of
the bead process introduced in \cite{FFN08}. But note that the interpretations of the variables are different: in \cite{FFN08} $u_j^{(l)}$ is the position of the particle $j$ on line $l$ while here
$u_j^{(l)}$ is a (transformed) height at a lattice point as specified by
(\ref{uH}). 

Also given in \cite{FFN08} is a discrete version of (\ref{9}), and furthermore summation
methods are presented to compute marginal PDFs for a single line. In
\cite[Prop.~10.2.3]{Fo10} this working is adopted to the present continuous setting.
In particular, it is shown \cite[Eq.~(10.66)]{Fo10} that integrating over lines $l=1,2,\dots,$ with configuration
$\{y_j^{(N+1)} \}_{j=1,\dots,N+1}$ gives
\begin{equation}\label{10}
\prod_{s=1}^N {1 \over s!} \prod_{1 \le j < k \le N + 1}
(y_j^{(N+1)} - y_k^{(N+1)}).
\end{equation}
By symmetry, this same expression results by integrating over lines $l=2N, 2N - 1,\dots,
N + 1$ (in that order) with configuration $\{y_j^{(N+1)}\}_{j=1,\dots,N+1}$ on line $N+1$.
Hence
\begin{equation}\label{C}
C = \Big ( \prod_{s=1}^N {1 \over s!} \Big )^2
\int_{\tilde{Y}} \prod_{1 \le j < k \le N + 1} (y_j^{(N+1)} - y_k^{(N+1)} )^2 \, d \vec{y}^{(N+1)}
\end{equation}
where $\tilde{Y}$ denotes the region
$$
1 > y_1^{(N+1)} > y_2^{(N+1)} > \cdots > y_{N+1}^{(N+1)} > 0.
$$

The integrand in (\ref{C}) is symmetric, so the region $\tilde{Y}$ can be replaced by
$[0,1]^{N+1}$ provided we divide by $(N+1)!$. Furthermore,
$$
\int_{[0,1]^{N+1}} \prod_{1 \le j < k \le N + 1}
( y_j^{(N+1)} - y_k^{(N+1)} )^2 \, d \vec{y}^{(N+1)} =
\prod_{j=0}^N {\Gamma^2(1 + j)\Gamma(2 + j) \over
\Gamma (N + j + 1) }
$$
(this is a special case of the Selberg integral (see e.g.~\cite[Eq.~(4.3)]{Fo10}), or alternatively can be derived using knowledge of the normalization of single variable Jacobi polynomials \cite[Eq.~(5.74)]{Fo10})
and (\ref{C}) results. \hfill $\square$

\section{Uniformly distributed height variables}
\setcounter{equation}{0}
Suppose the height variables are sampled from the uniform distribution on $[0,1]$.
It was remarked during the proof of Proposition \ref{p5} that another
interpretation of the joint PDF (\ref{8}) for the heights in the directed solid-on-solid model 
is then as a joint PDF for interlaced
particles on $2N-1$ parallel line segments $[0,1]$, with $N(l)$ particles on line $l$.
It was furthermore remarked that this latter model is the so called finitized bead process, studied in  \cite{FFN08},
or more precisely the special case 
\begin{equation}\label{pq}
p=N, \qquad q = N
\end{equation}
 of the finitized bead process
(general $p$ and $q$ would correspond to the height model of Section 2 being defined
in a rectangle rather than square). 

Known results for the bead model can then be used to deduce properties of the
directed solid-on-solid model. Of particular interest for present purposes are the facts that
configurations of the bead process can be sampled by generating eigenvalues of a random matrix process, and
that the correlation functions have an explicit determinantal form. We will revise these
results separately, and discuss their consequence in relation to the directed solid-on-solid model.

\subsection{Sampling using random matrices}
What does a typical configuration of the directed solid-on-solid model look like?
Section 4 of \cite{FFN08} details how (a generalization) of 
the PDF (\ref{8}) in the case that the heights are sampled from the uniform distribution
$U_{[0,1]}$ can be obtained as the joint eigenvalue PDF of a sequence of
random matrices. For a given matrix $A$, the construction is based on random corank 1 projections,
$M = \Pi A \Pi$. Here $\Pi = \mathbb I - \vec{x} \vec{x}^\dagger$, where $x$ is a normalized
complex Gaussian vector of the same number of rows as $A$. The effect of the
random corank 1 projection is to  reduce by one the multiplicity of any degenerate
eigenvalues in $A$, and also to create a new zero eigenvalue.

The first step is to form $M_1 =  \Pi A_1 \Pi$, where $A_1 = {\rm diag}((0)^N,(1)^N)$ 
(the notation $(k)^N$ is used to denote $k$ repeated $N$ times). The
eigenvalue of $M_1$ then corresponds to the height $h_{(1)}^{(1)}$. Next, for
$r=2,\dots,N$, inductively generate $\{ h_i^{(r)}\}_{i=1,\dots,r}$ as the eigenvalues different
from 0 and 1 of the matrix $M_r := \Pi A_r \Pi$, where
$$
A_r = {\rm diag} \Big ( (0)^{N-r+1}, h_1^{(r-1)},\dots,h_{r-1}^{(r-1)},(1)^{N-r+1} \Big ).
$$
And after this, for $r=1,\dots,N-1$, inductively generate $\{h_i^{(p+r)} \}_{i=1,\dots,N-r}$
$(r=1,\dots,N-1)$ as the eigenvalues different from 0 of
$M_{N+r} = \Pi A_{N+r} \Pi$ where
$$
A_{N+r} = {\rm diag} \Big ( h_1^{(N+r-1)},\dots, h_N^{(N+r-1)} \Big ).
$$

Crucial to the practical implementation of this construction is the fact that if
$$
A = {\rm diag} \Big ( (a_1)^{s_1},(a_2)^{s_2},\dots, (a_n)^{s_n} \Big ),
$$
the $n-1$ eigenvalues different from $\{a_i\}_{i=1,\dots,n}$ and $0$ occur at the
zeros of the random rational function
$$
\sum_{i=1}^n {q_i \over x - a_i }
$$
where $(q_1,\dots,q_n)$ has the Dirichlet distribution D$[s_1,\dots,s_n]$
\cite{FR02b}. 

\begin{figure}[t]\label{f1}
\begin{center}
\includegraphics[scale=0.8]{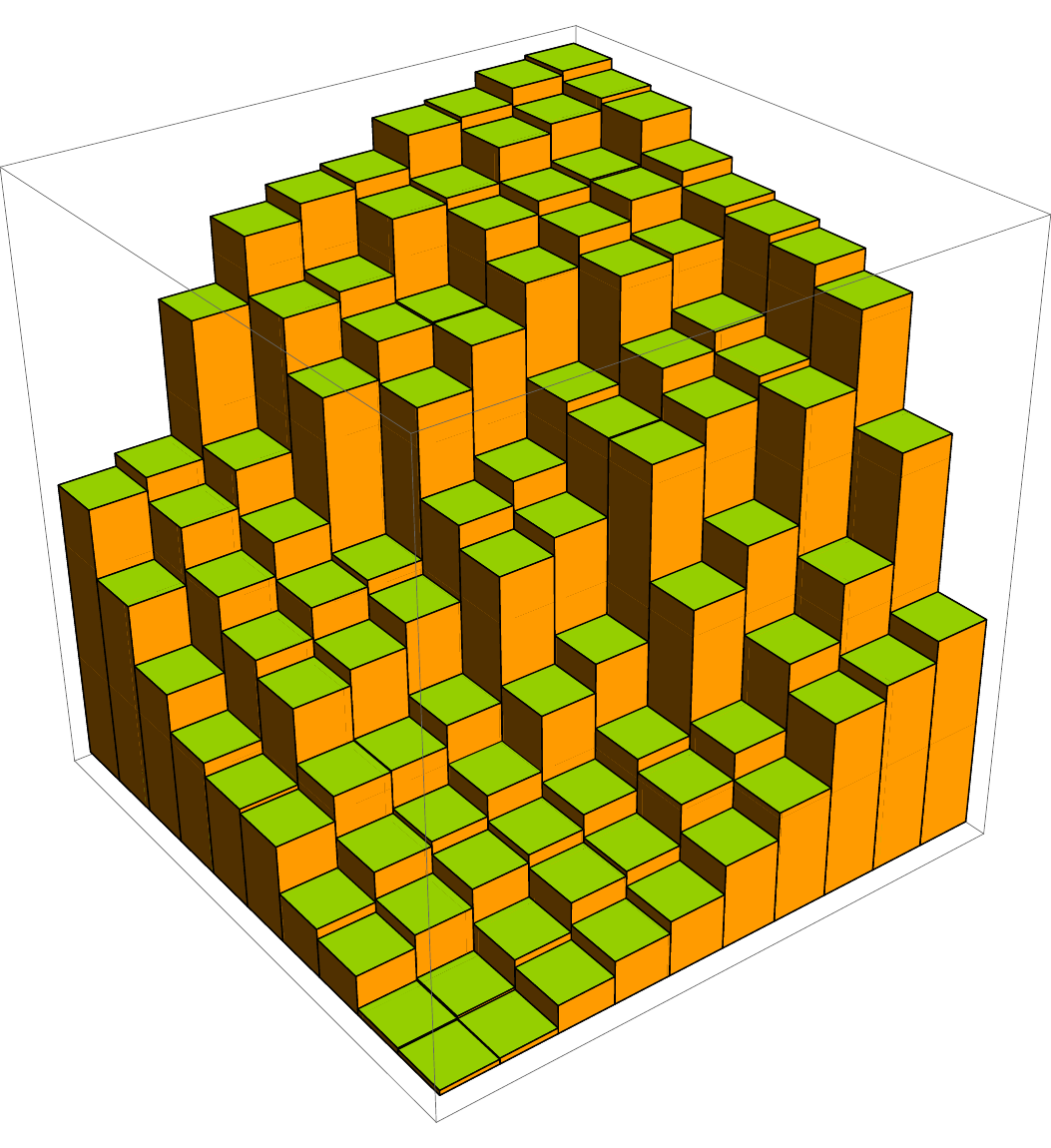}
\end{center}
\caption{A typical configuration of the directed solid-on-solid model on a
$10 \times 10$ grid, with the heights scaled by a factor of 10. This is constructed using
eigenvalues of a particular sequence of random matrices as described in the text. }
\end{figure} 

We have made use of the above theory to generate a typical configuration of the 
directed solid-on-solid
model in the case $p=q=10$, which is displayed graphically in Figure \ref{f1}.
We remark that if instead of sampling the individual heights
from the continuous uniform distribution, they were sampled instead from the discrete
uniform distribution on $[0,c]$ say ($c \in \mathbb Z^+$), then our directed solid-on-solid
model would be equivalent to boxed plane partitions, of box size $N \times N \times c$.
There is no longer a random matrix approach to the sampling of a typical configuration,
but we point out that one recently introduced method \cite{BG09} (see also
\cite{BF08} and \cite{BGR10}) has the interpretation as defining a dynamics on the underlying
interlaced particle configurations.

\subsection{Correlations and distribution functions}\label{sC}

Let $(l,x)$ refer to the position $x$ on line $l$ of the bead process. The correlation
function $\rho_{(k)}(X)$ for the configuration $X = \cup_{i=1}^s \{ l_i, x_1^{(l_i)},\dots,x_{r_i}^{(l_i)} \}$, 
where each $x^{(l_i)}$ is on line $l_i$ and $k = \sum_{i=1}^s r_i$, defined
as the average
\begin{equation}\label{3.10'}
\rho_{(k)}(X) = \Big \langle \prod_{i=1}^s  \sum_{m_1 \ne \cdots \ne m_{r_i}}^{N(l_i)}  \prod_{q=1}^{r_i}
\delta( x_q^{(l_i)}- y_{m_q}^{(l_i)}) \Big \rangle
\end{equation}
and thus requiring that  $k$ of the particles be fixed at positions $X$,
 has been
computed in \cite[Prop.~5.1]{FFN08}. The correlations are given
 in terms of the rescaled Jacobi polynomials
\begin{equation}\label{Jr}
\tilde{P}_n^{(a,b)}(x) := P_n^{(a,b)}(1-2x),
\end{equation}
which satisfy the orthogonality
\begin{equation}\label{Jorth}
\int_0^1 x^a (1 - x)^b \tilde{P}_j^{(a,b)}(x)  \tilde{P}_j^{(a,b)}(x)  \, dx =
{\mathcal N}_j^{(a,b)} \delta_{j,k}
\end{equation}
where
\begin{equation}\label{Jorth1}
{\mathcal N}_n^{(a,b)} = {1 \over 2n + a + b + 1}
{(n+a)! (n + b)! \over n! (n + a + b)!}.
\end{equation}
We also require the quantities
\begin{align}
&a_l(x) = \left \{ \begin{array}{ll} (-x)^{N-l} (1 - x)^{N - l}, & 1 \le l \le N \\
1, & N + 1 \le l \le 2 N -1 \end{array} \right. \label{A} \\
& b_l(x) = \left \{ \begin{array}{ll} (-1)^{N-l} , & 1 \le l \le N \\
x^{l-N} (1 - x)^{l - N}, & N + 1 \le l \le 2 N -1 \end{array} \right. \label{B} \\
& {\mathcal C}_j^{(l)} =
 \displaystyle {(l-j)! \over (N-j)!}, \quad 1 \le l \le 2N -1
\label{Ca} \\
& Q_j^{(l)}(x) =  \left \{ \begin{array}{ll} \tilde{P}_{l-j}^{(N-l,N-l)}(x), & 1 \le l \le N, \\
 \tilde{P}_{2N +1 - l-j}^{(l-N,l-N)}(x), & N+1 \le l \le 2N  - 1 \end{array}  \right. \label{Q} \\
 & {\mathcal N}_j^{(l)}(x) =
 \left \{  \begin{array}{ll}  {\mathcal N}_{l-j}^{(N-l,N-l)}, & 1 \le l \le N \\
 {\mathcal N}_{2N  -l-j}^{(l-N,l-N)}, & N+1 \le l \le 2N -1 \end{array}  \right.   \label{N} \\
 & \alpha(l,l') = {\rm min} \{2N+1-l,l',N \}.
 \end{align}
 
 In terms of this notation Proposition 5.1 of \cite{FFN08} gives
\begin{equation}\label{rK}
\rho_{(k)}(X) = \det [ K(l_i,x_i;l_j,x_j) ]_{i,j=1,\dots,k}
\end{equation}
where
\begin{equation}\label{rK1} 
K(s,u;t,v) =
\left \{  \begin{array}{ll}  \displaystyle a_s(u) b_t(v) \sum_{l=1}^{\alpha(s,t)}
{ {\mathcal C}_l^{(s)} \over {\mathcal C}_l^{(t)}}
{Q_l^{(s)}(u) Q_l^{(t)}(v) \over \mathcal N_l^{(t)}}, & s \ge t \\
\displaystyle - a_s(u) b_t(v) \sum_{l=- \infty }^0
{ {\mathcal C}_l^{(s)} \over {\mathcal C}_l^{(t)}}
{Q_l^{(s)}(u) Q_l^{(t)}(v) \over \mathcal N_l^{(t)}}, & s < t    \end{array}  \right.
\end{equation}

To make use of knowledge of this bead process correlation function in the context of the
solid-on-solid model, we must use it to specify the distribution of the position of specific particles.
In fact it is sufficient to specify $E_0(\{l_i;(u_i,1) \}_{i=1,\dots,n})$ --- the probability 
there are no particles in the interval $(u_i,1)$ of line
$l_i$ $(i=1,\dots,n)$. Thus the PDF
$p^{\rm max}(\{l_i,u_i\}_{i=1,\dots,n})$ for the position of the particles with maximum displacements on lines $l_i$,
$i=1,\dots,n$, then follows by partial differentiation according to
\begin{equation}\label{Kt2}
p^{\rm max}(\{l_i,u_i\}_{i=1,\dots,n}) = {\partial^n \over \partial u_1 \cdots \partial u_n}
E_0(\{l_i;(u_i,1) \}_{i=1,\dots,n}).
\end{equation}
The crucial point from the viewpoint of the corresponding height model is that this PDF is
identical to the PDF for the maximum heights on the same lines $i=1,\dots,n$.

We are thus faced with the task of expressing $E_0(\{l_i;(u_i,1) \}_{i=1,\dots,n})$ in terms of correlation functions,
which is in fact a standard exercise \cite[\S 8.1]{Fo10}. First note that according to the definition
$$
E_0(\{l_i;(u_i,1) \}_{i=1,\dots,n}) = \Big \langle \prod_{i=1}^n \prod_{l=1}^{N(l_i)}(1 - \chi_{y_l^{(l_i)} \in (u_i,1)})
\Big \rangle.
$$
Expanding the double product and recalling the definition (\ref{3.10'}) of the correlation function then shows
\begin{eqnarray}\label{E4}
&& E_0(\{l_i,(u_i,1)_{i=1,\dots,n})  = \sum_{m_1 = 0}^{N(l_1)} \cdots  \sum_{m_n = 0}^{N(l_n)} 
{(-1)^{m_1+\cdots + m_n} \over m_1! \cdots m_n!}
\nonumber \\
&& \quad \times \int_{u_1}^1 dy_1^{(l_1)} \dots \int_{u_1}^1 dy_{m_1}^{(l_1)}
\cdots  \int_{u_n}^1 dy_1^{(l_n)} \dots  \int_{u_n}^1 dy_{m_n}^{(l_n)} \,
\rho_{(\sum_{i=1}^n m_i)}(\cup_{i=1}^n \cup_{j=1}^{m_i} \{ (l_i, y_j^{(l_i)}) \} ), \qquad
\end{eqnarray}
where the term $m_1 = \cdots = m_n = 0$ is taken to equal unity. 

This is a general formula valid for any two-dimensional particle
system confined to $2N-1$  parallel lines, with $N(l)$ particles on line $l$ ($l=1,\dots,2N-1$).
It is furthermore the case that when the $k$-point ($k= \sum_{i=1}^n m_i$) correlation function $\rho_{(k)}(Y)$
has a determinantal form (\ref{rK}), the multiple sum (\ref{E4}) can be summed \cite[\S 9.1]{Fo10}. 
This can be done by defining the $n \times n$ matrix Fredholm integral operator
$K(\{(l_i;(u_i,1))\}_{i=1,\dots,n})$ with kernel
\begin{equation}\label{Kt}
\tilde{K}(x,y; \{l_i;(u_i,1) \}_{i=1,\dots,n}) =  \Big [ \chi_{x \in (u_i,1)} K(x,l_i;y,l_j) \chi_{y \in (u_j,1)} \Big ]_{i,j=1,\dots,n}.
\end{equation}
We then have that 
\begin{equation}\label{Kt1}
E_0(\{l_i;(u_i,1) \}_{i=1,\dots,n}) = \det \Big ( {\bf 1} -{K}(\{l_i;(u_i,1) \}_{i=1,\dots,n}) \Big ),
\end{equation}
where the meaning of the determinant can be taken as the product over the eigenvalues of the operator.

\section{Scaled limits for uniformly distributed heights}
\setcounter{equation}{0}
\subsection{The limiting shape}
In the large $N$ limit the global density on line of the finitized bead process has been computed
in \cite{FFN08}. With the parameters as in \ref{pq}, and the label $j$ of each line scaled
$j \mapsto j/N =: S$ and thus $0 < S < 2$, the support of the density was found to be the
interval
\begin{equation}\label{c1}
\Big [ {1 \over 2} - {1 \over 2} \sqrt{S(2-S)}, {1 \over 2} + {1 \over 2} \sqrt{S(2-S)} \Big ] =:
[ c_S, d_S].
\end{equation}
And after dividing by $N$, the explicit functional form of the density was shown to equal
\begin{equation}\label{c2}
\rho_{(1)}(y,S) = {1 \over \pi} {\sqrt{(d_S - y)(y - c_S)} \over y(1 - y)}.
\end{equation}

To interpret these results in terms of the directed solid-on-solid model with heights
sampled from $U_{[0,1]}$, we first agree to scale the $N \times N$ integer grid by $1/N$ so that
it is an $N \times N$ grid within the unit square $[0,1]^2$. It then follows immediately from (\ref{c1})
that the limiting height profiles, $h(x,y)$ say, along $y=1$, $x=1$, $y=0$ and $x=0$
respectively are
\begin{align}\label{c3}
& h(x,1) = {1 \over 2} ( 1 + \sqrt{x(2 - x)} ) \nonumber \\
& h(1,y) = {1 \over 2} ( 1 + \sqrt{y(2 - y)} ) \nonumber \\
& h(x,0) = {1 \over 2} ( 1 - \sqrt{1 - x^2} ) \nonumber \\
& h(0,y) = {1 \over 2} ( 1 - \sqrt{1 - y^2} ) 
\end{align}
We remark that (\ref{c3}) exhibits the general symmetry $h(x,y) = h(y,x)$, which in turn follows
from the symmetry with respect to the $x$ and $y$ directions of the rule for the interlacing of
heights in the definition of the model.

To specify the heights at other positions of the unit square we need to work in the
coordinates which relate directly to those used for the scaled bead process.
The variable $S$ used in (\ref{c1}), for $0\le S\le1$, then identifies the line segment in
$[0,1]^2$ starting at $(0,1-S)$ and finishing at $(S,1)$. Let $t$, $0 < t < 1$ denote the
scaled position along this line segment so that
\begin{equation}\label{xyS}
(x,y) = (t S, 1 + S(t-1)).
\end{equation}

We seek the height $h(S,t)$ at position $t$ along line $S$. A little thought
(cf.~\cite[eq.~(3.5)]{FS03}) shows that $h(S,t)$ is characterized by the equation
\begin{equation}\label{4c}
t S = \int_{c_S}^{h(S,t)} \rho_{(1)}(u,s) \, du.
\end{equation}
Thus in the bead model picture, the RHS of (\ref{4c}) gives the expected number of particles
from the start of the line segment, to position $h(S,t)$ along the segment. The LHS says this
number of particles is equal to $tS$. Hence particle number $tS$ along this segment is
expected to be at position $h(S,t)$ on the line. In the solid-on-solid picture, the expected
position of a particular numbered particle is the expected 
height at the point in the square corresponding to the  numbering of the particle.

We note from (\ref{c2}) and (\ref{c1}) that $\rho_{(1)}(y,S)$ is symmetrical about $y=1/2$.
It follows that
$$
\int_{c_S}^{1/2}  \rho_{(1)}(u,s) \, du = {1 \over 2} \int_{c_S}^{d_S}  \rho_{(1)}(u,s) \, du =
{S \over 2}, \qquad 0 \le S  \le 1
$$
and consequently
$$
h(S, {1 \over 2} ) = {1 \over 2}, \qquad 0 \le  S \le 1
$$
or equivalently in terms of $xy$-coordinates,
\begin{equation}\label{xy2}
h(x,1-x) = {1 \over 2}, \qquad 0 \le x \le 1.
\end{equation}
This supplements the exact profiles (\ref{c3}).

\begin{figure}[t]\label{f2}
\begin{center}
\includegraphics[scale=0.8]{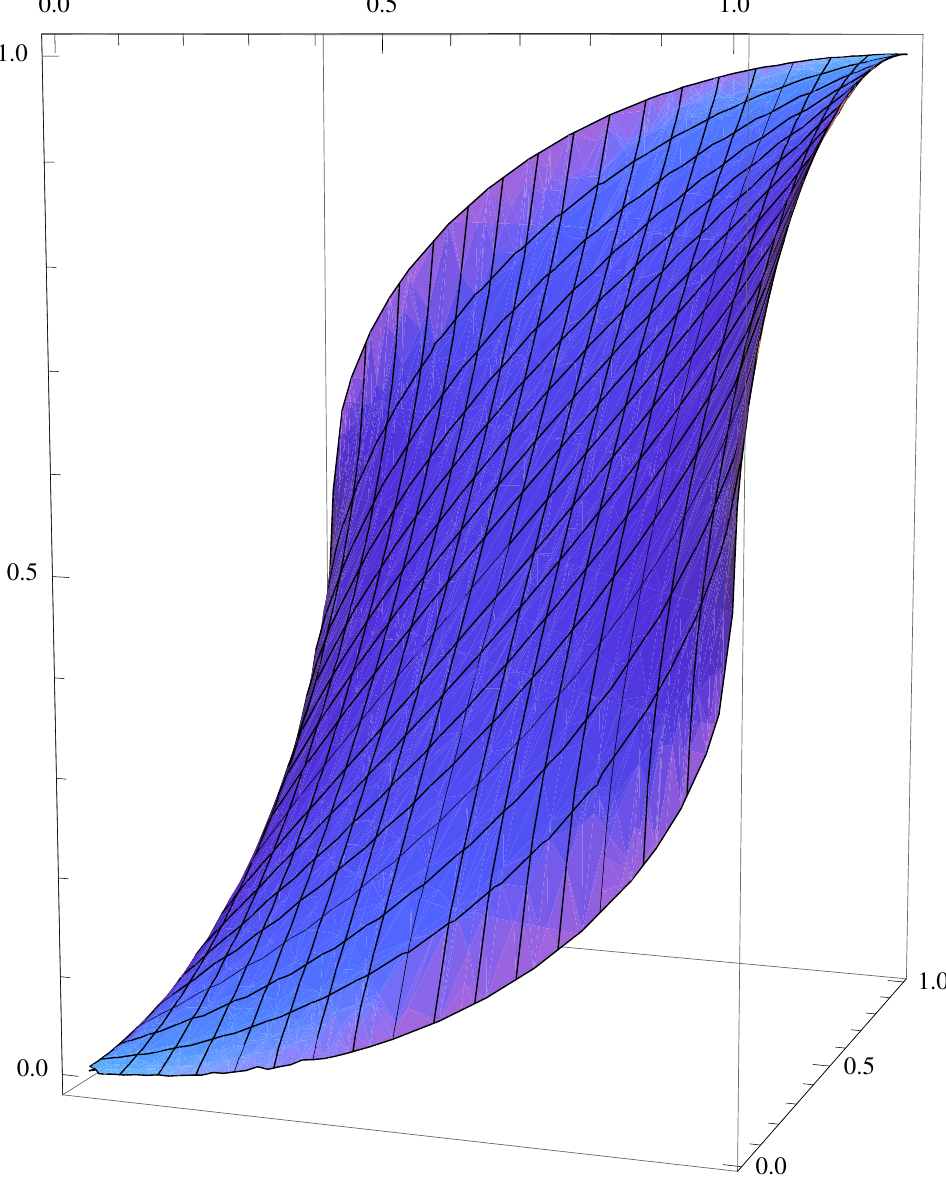}
\end{center}
\caption{A plot of the limiting surface of the directed solid-on-solid model as calculated using
(\ref{4d}) and (\ref{bbv}).}
\end{figure}

Along other line segments of the unit square, we have to make do with (\ref{4c}),
although the integral can be evaluated explicitly. Thus recalling (\ref{c1}) and (\ref{c2}),
use of computer algebra allows us to conclude
\begin{equation}\label{4d}
t S =   {1 \over  \pi} \Big (  \arcsin v +\pi/2 + \sqrt{1 - \tilde{d}^2}
(\arctan \Big ( {\sqrt{1 - \tilde{d}^2} \over \sqrt{1 - v^2}} v \Big ) + \pi/2)\Big )
\end{equation}
where
\begin{equation}\label{4e}
v := (2 h(S,t) - 1)/ \tilde{d}, \qquad \tilde{d} = \sqrt{S(2-S)}.
\end{equation}
For practical determination of the height profile, we thus choose $(x,y)$ within the unit square,
then determine $S$ and $t$ according to
\begin{equation}\label{bbv}
S = 1 + x - y, \qquad t = x/(1 + x - y)
\end{equation}
as implied by (\ref{xyS}). Substituting these values in (\ref{4d}) we obtain an equation for
$h$, which is solved using a root finding routine.
A graph of the resulting shape is plotted
in Figure \ref{f2}.

\subsection{Scaled correlation of heights along $y=1$}
The expected value of the heights along $y=1$ in the limiting solid-on-solid model
is the first entry in (\ref{c3}). We seek a scaled limit of the joint distribution function
(\ref{Kt2}) for $k$ heights along this line, which according to (\ref{Kt1})
requires computing the large $N$ form of  (\ref{Kt}). In the limiting procedure
$x$ and $y$ must first undergo a linear change of scale. Thus after
introducing the scaled line number $S_j$, for $x_j$ corresponding to this line we must write
\begin{equation}\label{xs}
x_j = x_0(S_j) + X_j \sigma N^{-2/3}
\end{equation}
where $ x_0(S) := {1 \over 2} ( 1 + \sqrt{S(2 - S)})$, $\sigma$ is independent of
$N$ and can be chosen at our convenience, and the factor $N^{-2/3}$ is chosen
so that in the variable $X$ the spacing between the large heights on line $S$ is
${\rm O}(1)$. We must also choose the spacing between scaled lines to scale with $N$.
Thus we write
\begin{equation}\label{xsS}
S_j = S^* + t_j \tau N^{-1/3}
\end{equation}
where $S^*$ is the limiting line on the scale of a division by $N$, and $\tau_{S^*}$ is independent of $N$ to be
chosen for convenience. The motivation for the choice (\ref{xsS}) is that the scaled
correlations are the ${\rm O}(1)$ in the variables $\{t_j\}$.

Since from \S \ref{sC} the correlations are given in terms of Jacobi polynomials
(\ref{Jr}), we
require a uniform asymptotic expansion appropriate for the scaling (\ref{xs}). The relevant
such expansion has recently been given by Johnstone \cite{Jo08}: it applies when the
Jacobi polynomial parameters $a,b$ increase with $n$, and when $x$ is centred about the
largest (or by replacing the roles of $a,b$, the smallest) zero, with a further scaling chosen so that the spacing between zeros is of
order unity. For our purposes, due to the relation (\ref{Jr}), we formulate the results
of \cite{Jo08} about the smallest zero of $P_n^{(a,b)}(x)$, and thus the largest zero of
$\tilde{P}_n^{(a,b)}(x)$.

\begin{proposition}\label{p41}
Define variables $\kappa_n$, $\psi$ and $\gamma$ by
\begin{equation}\label{3.147}
\kappa_n = 2n + a + b + 1, \quad \cos \psi = {a - b \over \kappa_n}, \quad
\cos \gamma = {a + b \over \kappa_n},
\end{equation}
and using these variables define
\begin{equation}\label{3.148}
M_n =  \cos ( \psi + \gamma), \qquad \sigma_n^3 = {2 \sin^4(\psi + \gamma) \over
\kappa_n^2 \sin \psi \sin \gamma}.
\end{equation}
Then, for $x = M_n - \sigma_n X$, one has the uniform asymptotic expansion
\begin{equation}\label{3.149}
P_n^{(a,b)}(x) = \sqrt{ {\kappa_n \sigma_n 2^{a+b+1}{\mathcal N}_n^{(a,b)} \over
(1 - x)^a (1 + x)^b (1 - x^2)}} \bigg ( {\rm Ai} (X) + {\rm O}(N^{-2/3})
\left \{ \begin{array}{ll} e^{- X/2}, & X > 0 \\
1, & X < 0 \end{array} \right. \bigg ).
\end{equation}
Equivalenty, in terms of the rescaled Jacobi polynomials (\ref{Jr}),
for $x = (1 - M_n)/2 + \sigma_n X/2$,
\begin{equation}\label{3.149}
\tilde{P}_n^{(a,b)}(x) = \sqrt{ {\kappa_n \sigma_n {\mathcal N}_n^{(a,b)} \over
2 x^{a+1} (1 - x)^{b+1} }} \bigg ( {\rm Ai} (X) + {\rm O}(N^{-2/3})
\left \{ \begin{array}{ll} e^{- X/2}, & X > 0 \\
1, & X < 0 \end{array} \right.  \bigg ).
\end{equation}

\end{proposition}

According to (\ref{Q}), with $l$ scaling with $N$ such that $ \tilde{l} := l/N$ is fixed,
we have $a =  b= N - l$,
and thus $\cos \psi = 0$,
$ \lim_{l \to \infty} \cos \gamma = 1 -  \tilde{l} $, which together imply
\begin{equation}\label{ML}
{1 \over 2} - {1 \over 2} \lim_{l \to \infty} M_l = {1 \over 2} +
{1 \over 2} \sqrt{1 - (1 - \tilde{l})^2}.
\end{equation}
With $\tilde{l}$ identified as $S$, this is precisely $x_0(S)$ in (\ref{xs}). We want to use
Proposition \ref{p41} to show that with an appropriate choice of scaled variables as implied by
(\ref{xs}) and (\ref{xsS}),
the summations  \eqref{rK1}  defining the correlation kernel
$K$ are slowing varying in the summation index $l$ and as $N \to \infty$ are given by
explicit Riemann integrals. The latter we will recognise as specifying the well known Airy process
 from random matrix theory  \cite{Ma94,FNH99}. In addition, for purposes of application to the
 scaled limit of the probability $E_0$ we need to demonstrate that the error term
 is integrable to the right of the soft edge.

\begin{proposition}\label{Pw}
Let
\begin{eqnarray}\label{xXyY}
x = x_0(S_x) + X \sigma N^{-2/3}, & & y = x_0(S_y) + Y \sigma N^{-2/3}, \nonumber \\
S_x = S + s \tau N^{-1/3}, & & S_y = S+t \tau N^{-1/3},
\end{eqnarray}
with $S  \ne 1$, $x_0(S)$  as specified below (\ref{xs}) and $\sigma$, $\tau$ such that
\begin{equation}\label{tau1}
\sigma^3 = \frac{(1-S)^4}{16\sqrt{S(2-S)}}, \qquad 
\tau = \frac{(1-S)^2\sqrt{S(2-S)}}{2\sigma}.
\end{equation}
With $K$ as in \eqref{rK1}, define $\bar{K}$ by
\begin{equation}
\bar{K} (X, s ; Y, t) = \lim_{N\rightarrow \infty} \sigma N^{-2/3} K (x, NS_x; y, NS_y).
\end{equation}
We then have
\begin{align}
\bar{K} (X, s ; Y, t) &= \left \{ \begin{array}{ll} \displaystyle \frac{F(X,s)}{F(Y,t)} \int_0^{\infty} e^{(t-s)u}
 {\rm Ai} \left(u + X\right){\rm Ai} \left(u + Y\right) du, & s \ge t \\
 \displaystyle - \frac{F(X,s)}{F(Y,t)} \int_{-\infty}^0 e^{(t-s)u}
 {\rm Ai} \left(u + X\right){\rm Ai} \left(u + Y\right) du, & s < t \end{array} \right. \nonumber \\
 & =:  \frac{F(X,s)}{F(Y,t)} K^{\rm AiryP} (X, s ; Y, t) 
\end{align}
and consequently
\begin{align}\label{AC}
\bar{\rho}_{(k)}(\{(X_j,s_j) \}_{j=1,\dots,k}) & := \lim_{N \to \infty} ( \sigma N^{-2/3})^k \rho_{(k)}((x_j,NS_{x_j})_{j=1,\dots,k}) \nonumber \\
& = \det [  K^{\rm AiryP} (X_j, s_j ; X_l, s_l) ]_{j,l=1,\dots,k}.
\end{align}
Moreover,
\begin{align}\label{AC1}
&  \lim_{N \to \infty} ( \sigma N^{-2/3})^k \int_{y_1}^1 dx_1 \cdots  \int_{y_k}^1 dx_k \,
\rho_{(k)}(\{ (x_j,N S_{x_j})\}_{j=1,\dots,k}) \nonumber  \\ & \quad =
\int_{Y_1}^\infty dX_1 \cdots \int_{Y_k}^\infty dX_k \,
\bar{\rho}_{(k)} (\{ (X_j,s_j) \}_{j=1,\dots,k}).
\end{align}
\end{proposition}

Proof. \quad Let us introduce $l_x$ according to $S_x = l_x/N$. Substituting $x$ as given in \eqref{xXyY} into the uniform asymptotic expansion for the Jacobi polynomial  \eqref{3.149}, with $n = l_x-j $, $a=b=N-l_x$, gives
\begin{equation}\label{3.150}
P_{l_x-j}^{(N-l_x,N-l_x)}(x) = \sqrt{ {\kappa_{l_x-j} \sigma_{l_x-j} {\mathcal N}_{l_x-j}^{(N-l_x,N-l_x)} \over
2x^{N-l_x+1} (1 - x)^{N-l_x+1} }} \bigg ( {\rm Ai} (X^*) + {\rm O}(N^{-2/3})
\left \{ \begin{array}{ll} e^{- X^*/2}, & X^* > 0 \\
1, & X^* < 0 \end{array} \right.  \bigg ),
\end{equation}
where, in the notation of (\ref{3.148}), $X^*$ is given by
\begin{equation}\label{Xs}
X^* =  \frac{2}{\sigma_{l_x-j}} \left(\frac{M_{l_x-j}-1}{2} +x\right).
\end{equation}
Recalling the definition \eqref{rK1} and using the expansion \eqref{3.150}, $K(x, l_x; y , l_y)$ is given by a summation over $j$ of 
 \begin{align}\label{bigeq}
\frac{(-1)^{l_x-l_y}}{2} \sqrt{\frac{(x(1-x))^{N-l_x-1}}{(y(1-y))^{N-l_y+1}}} \frac{(l_x-j)!}{(l_y-j)!} & \sqrt{\frac{{\mathcal N}_{l_x-j}^{(N-l_x,N-l_x)}}{{\mathcal N}_{l_y-j}^{(N-l_y,N-l_y)}}} \sqrt{\kappa_{l_x-j} \kappa_{l_y-j}  \sigma_{l_x-j} \sigma_{l_y-j}}   \\
\times \bigg ( {\rm Ai} (X^*)  {\rm Ai} (Y^*) + {\rm O}(N^{-2/3})
& \bigg (  \left \{ \begin{array}{ll} e^{- X^*/2}, & X^* > 0 \\
1, & X^* < 0 \end{array} \right.  \bigg )
 \bigg ( \left \{ \begin{array}{ll} e^{- Y^*/2}, & Y^* > 0 \\
1, & Y^* < 0 \end{array} \right.  \bigg ) \bigg ).
\nonumber \end{align}

In fact the summand (\ref{bigeq}) is a slowly varying function of $j$. To see this 
we first note from (\ref{3.148}) that
\begin{equation}
M_{l_x-j} = -\sqrt{S_x(2-S_x)} - \frac{j}{N} \frac{(S_x-1)^2}{S_x(2-S_x)} + {\rm max} \Big ({\rm O} (N^{-1}), {\rm O} (j^2/N^2) \Big )
\end{equation}
and $\sigma_{l_x-j} = 2\sigma N^{-2/3} + {\rm O} (N^{-1})$.  This substituted in (\ref{Xs}) suggests  we introduce a continuous summation label $w$ by
$j=wN^{1/3}$ to obtain the large $N$ form 
\begin{equation}
X^* = -w\frac{(S-1)^2}{2\sigma \sqrt{S(2-S)}} + X + {\rm O} (N^{-1/3}).
\end{equation}
Furthermore, in terms of $w$ use of Stirling's formula shows that the first line of \eqref{bigeq} (the factors before the Airy functions) reads
\begin{equation}
\frac{F(X,s)}{F(Y,t)}  \left(\frac{8N^{1/3}\sigma}{(S-1)^2}\right) \exp\left(\frac{(s-t)\tau w}{S(2-S)}\right)\left(1 + {\rm O}(N^{-1/3})\right)
\end{equation}
with the error term uniform in $w$ and where
\begin{equation}
F(X,s) = (-1)^{l_x}[x(1-x)]^{(N-l_x-1)/2}N^{s\tau N^{2/3}} \left(S(2-S)\right)^{s\tau N^{2/3}/2}e^{\frac{-s^2\tau^2(1-S)}{S(2-S)}N^{1/3} +  \frac{s^3\tau^3(2-2S+S^2)}.{3S^2(2-S)^2} }
\end{equation}
Thus, to leading order the summand depends only on the continuous summation label $w$, and thus can be converted to a Reimann integral to give
\begin{align}\label{abc}
 \sigma N^{-2/3} K (x, NS_x; y, NS_y)
 = & \frac{F(X,s)}{F(Y,t)}    \left(\frac{8\sigma^2}{(S-1)^2}\right) \int_0^{\infty}\exp\left(\frac{(s-t)\tau w}{S(2-S)}\right ) \nonumber\\
& \quad \times {\rm Ai} \left(-w\frac{(S-1)^2}{2\sigma \sqrt{S(2-S)}} + X\right)  {\rm Ai} \left(-w\frac{(S-1)^2}{2\sigma \sqrt{S(2-S)}} + Y\right) dw  \nonumber\\ 
&\qquad + {\rm O}(N^{-1/3}) {\rm O}(e^{-(X+Y)/2}).
\end{align}
A change of variables
\begin{equation}
u = \frac{-w(S-1)^2}{2 \sigma \sqrt{S(2-S)}}
\end{equation}
turns this into
\begin{align}
\lim_{N \to \infty} \sigma N^{-2/3} K (x, NS_x; y, NS_y)
=\frac{F(X,s)}{F(Y,t)}   \int_0^{\infty} \exp\left(\frac{(t-s)\tau 2 \sigma u}{\sqrt{S(2-S)}(1-S)^2}\right) \nonumber  \\
\times {\rm Ai} \left(u + X\right){\rm Ai} \left(u + Y\right) du
\end{align}
and the choice of $\tau$ as in (\ref{tau1})
completes the proof in the case $s \ge t$.  Furthermore, essentially the same working holds true for the remaining case $s < t$. 

With the form of $\bar{K}$ now established, the formula (\ref{AC}) for the
correlation function follows by noting that upon substitution in (\ref{rK}) the factor $  F(X_j,s_j) / F(X_l,s_l) $ cancels from the determinant.
Finally, the validity of (\ref{AC1}) follows from (\ref{rK}) combined with the error term in (\ref{abc}).
\hfill $\square$

We can make use of Proposition \ref{Pw} to compute the scaled limit of the probability $E_0$. Thus we know from
\cite{So01}, \cite{BF03} that the convergence of the integrals (\ref{AC1}) implies that the scaled limit can be applied term-by-term
in (\ref{E4}). Since the limiting correlations are determinantal, the resulting sum can again be summed by appealing to 
Fredholm integral operator theory  (see e.g.~\cite[\S 9.1]{Fo10}).

\begin{cor}
Let $v_j$ and  $V_j$ be related as implied by the first line in (\ref{xXyY}), and $s_j$ and $S_j$ be related as implied by the second line.
We have
\begin{equation}\label{Sb}
\lim_{N \to \infty} E_0(\{NS_i;(v_i,1) \}_{i=1,\dots,n}) =
\det \Big ( {\bf 1} - K^{\rm AiryP}( \{ s_i, (V_i,\infty) \}_{i=1,\dots,n}) \Big )
\end{equation}
where $K^{\rm AiryP}( \{ s_i, (V_i,\infty) \}_{i=1,\dots,n})$ is the $n \times n$ matrix Fredholm integal operator with kernel
\begin{equation}\label{Kta}
\tilde{K}^{\rm AiryP} (X,Y; \{s_i;(V_i,\infty) \}_{i=1,\dots,n}) =  \Big [ \chi_{X \in (V_i,\infty)} K^{\rm AiryP}(X,s_i;Y,s_j) \chi_{Y \in (V_j,\infty)} \Big ]_{i,j=1,\dots,n}.
\end{equation}
\end{cor}

The expression (\ref{Sb}) is precisely that for the cumulative distribution function of the scaled largest eigenvalue in the
Dyson Brownian motion model of complex Hermitian matrices \cite{PS01} (see \cite[Ch.~11]{Fo10} for an account of the
model). We remark that the statistical system defined by the scaling of the eigenvalues in the
Dyson Brownian motion model of complex Hermitian matrices about the largest eigenvalue is referred to as
the Airy process  \cite{PS01}. And as noted in the Introduction, the Airy process also occurs in random matrix theory
as the correlation functions for the GUE minor process at the soft edge \cite{FN08}.

\section{Universality}
\setcounter{equation}{0}
We know turn our attention to the case of a general absolutely continuous distribution $\mathcal D$ for the height
variables, characterized by the corresponding PDF $h(x)$. We have already seen that the change of variables (\ref{uH})
maps the joint PDF for configurations of interlaced heights sampled from $\mathcal D$ to the joint PDF for configurations of interlaced heights 
sampled from the particular case $\mathcal D = U_{[0,1]}$ --- the uniform distribution on $[0,1]$. And in the previous section we have identified as
the Airy process a
particular scaling limit of the joint distribution of the heights along $y=1$ in the case $\mathcal D = U_{[0,1]}$.
Our aim in this subsection is to show that the change of variables (\ref{uH}) implies that there is a choice of scaled variables  which also gives the Airy process as the scaling limit of the joint distribution of the heights
along $y=1$.

Let $N$ be large, and scale the vertical lines in the corresponding bead process so they are labelled by a continuous
parameter $S$ as specified at the beginning of \S 4.1. Again in the bead process picture, (\ref{c1}) gives the support of the
density in the case $\mathcal D = U_{[0,1]}$. It follows immediately from the change of variables (\ref{uH}) that the along line $S$ the
interval of support for a general absolutely continuous distribution with PDF $h$ is specified by the equations
\begin{equation}\label{chd}
{1 \over 2} - {1 \over 2} \sqrt{S(2-S)} = \int_0^{c_S(h)} h(t) \, dt, \qquad
{1 \over 2}  + {1 \over 2} \sqrt{S(2-S)} = \int_0^{d_S(h)} h(t) \, dt.
\end{equation}
We note that $d_S(h)$ must be finite for $S \ne 1$. Recalling (\ref{xs}) in the case $\mathcal D = U_{[0,1]}$ for a scaling of positions in the
bead model/ heights in the directed solid-on-solid model about the upper edge of support, we introduce the renormalized parameter $\tilde{\sigma}$
such that
\begin{equation}\label{dd}
d_S(1) + X_j \sigma N^{-2/3} = \int_0^{d_S(h) + X_j \tilde{\sigma} N^{-2/3}} h(t) \, dt.
\end{equation}
For $h(t)$ continuous we ca expand the RHS to leading order $N^{-2/3}$ and so deduce that to leading order in $N$,
\begin{equation}\label{shs}
\sigma = \tilde{\sigma} h(d_S(h)).
\end{equation}
Hence we conclude that for $h(t)$ continuous and non-zero at $t= d_S(h)$, the scaling of positions about the upper edge of the
support
\begin{equation}\label{shs1}
x = d_{S_x}(h) + X_j/(h(d_{S_x}(h)) N^{2/3})
\end{equation}
is consistent with the change of variables (\ref{uH}) required for the equality of joint probabilities (\ref{9}). Thus with
the change of variables (\ref{shs1}) the scaling limit of the joint distribution of heights in the general case 
must be the same as in the particular case $\mathcal D = U_{[0,1]}$, so giving our sought universality result.

\begin{cor}\label{cE}
Let $v_j$ and $V_j$ be related as for $x_j$ and $X_j$ in (\ref{shs1}). Let $s_j$ and $S_j$ be related as given in the second
line of (\ref{xXyY}), where again it is required $S \ne 1$. For the directed solid-on-solid model with heights sampled from a
general absolutely continuous distribution $\mathcal D$ with corresponding PDF $h(x)$, $h(x)$ itself continuous and assumed non-zero at 
the boundary reference point $x = d_S(h)$, the limit formula (\ref{Sb}) remains valid.
\end{cor}

The requirement that $S \ne 1$ is crucial ($S$ is the reference line in the bead process picture about which, on separations
of order $N^{2/3}$, the correlations between heights along $y=1$ are being measured). To see this, let us consider the fluctuation
of the height $x_{N,N}$ in the top right corner of our directed solid-on-solid model. In the bead process picture, corresponds to the
particle with the largest coordinate on the  line $S=1$. Now according to the definition of the model,
$x_{N,N}$ is the largest of the heights. Hence, for $x_{N,N}$ to be less than $X$ we must have that all the
$N(N+1)/2$ height variables must be less that $X$, and so
\begin{equation}\label{k1}
{\rm Pr} ( x_{N,N} < X) = \Big ( \int_0^X h(t) \, dt \Big )^{N(N+1)/2}
\end{equation}
(here the interlacing requirement plays no role). 

In the case $\mathcal D = U_{[0,1]}$ we see from this that 
\begin{equation}\label{k2}
\lim_{N \to \infty} {\rm Pr} ( x_{N,N} < 1 - X/N^2) = e^{-X/2}.
\end{equation}
On the other hand, for $\mathcal D$ the exponential distribution it follows from (\ref{k1}) that
\begin{equation}\label{k2}
\lim_{N \to \infty} {\rm Pr} ( x_{N,N} < X + 2 \log N) = e^{-e^{-X}}.
\end{equation}
Thus for the fluctuations of the maximum height, both the scale required to get a well defined limiting
distribution, and the limiting distribution itself, are dependent on $\mathcal D$, in contrast to the situation
exhibited in Corollary \ref{cE}.

\subsection*{Acknowledgements}
The work of the authors was supported by a Melbourne Postgraduate Research Award and the Australian
Research Council respectively. PJF thanks Eric Nordenstam for a discussion on this topic at the
MSRI random matrix semester, Fall 2010.


\providecommand{\bysame}{\leavevmode\hbox to3em{\hrulefill}\thinspace}
\providecommand{\MR}{\relax\ifhmode\unskip\space\fi MR }
\providecommand{\MRhref}[2]{%
  \href{http://www.ams.org/mathscinet-getitem?mr=#1}{#2}
}
\providecommand{\href}[2]{#2}

\end{document}